\documentstyle[12pt]{article}\textheight 230mm\textwidth 160mm
\newcommand{\bi}{\bibitem}
\newcommand{\be}{\begin{eqnarray}}
\newcommand{\ee}{\end{eqnarray}}
\newcommand{\nn}{\nonumber}

\begin{document}

\vspace{-1cm}
\hspace*{11cm}MPI-PhT/94-01\\
\hspace*{11.6cm}KANAZAWA-94-01\\
\hspace*{11.6cm}January 1994
\begin{center}
 {\bf\Large An analysis on the convergence of
equal-time commutators\\and\\ the closure of
the BRST algebra in Yang-Mills theories}
\end{center}
\vspace*{0.5cm}
\begin{center}{\sc\large Jisuke Kubo$\ ^{\dagger~*}$}
\end{center}
%\vspace*{0.2cm}
\begin{center}
{\em  Max-Planck-Institut f\"ur Physik\\
- Werner-Heisenberg-Institut -\\
P.O.Box 40 12 12, Munich (Fed. Rep. Germany)}
\end{center}
%\vspace*{0.5cm}
\noindent
{\sc\large Abstract}
\newline
\noindent
In renormalizable theories, we define
equal-time commutators (ETC'S) in terms of
the equal-time limit and investigate its convergence
in perturbation theory.
We find that the equal-time limit vanishes for amplitudes
with the effective dimension $d_{\em eff} \leq -2$ and is finite
for those with $d_{\em eff} =-1$ but without nontrivial
discontinuity. Otherwise we expect divergent equal-time limits.
We also find  that, if the ETC's involved in verifying an Jacobi
identity  exist, the identity is satisfied.
 Under these circumstances, we show
in the Yang-Mills theory that the ETC
of the $0$ component of the BRST current with each other
vanishes to all orders in perturbation theory
if the theory is free from the chiral anomaly, from which we conclude
that $[\, Q\,,\,Q\,]=0$,
where $Q$ is the BRST charge.
For the case that the chiral anomaly is not canceled, we use
various broken Ward identities to show that
$[\, Q\,,\,Q\,]$ is finite and
 $[\,Q\,,\,[\, Q\,,\,Q]\,]$ vanishes at
the one-loop level and that they start to diverge at the two-loop
level unless there is some unexpected cancellation mechanism
that improves the degree of convergence.
\vspace*{0.5cm}
\footnoterule
\vspace*{2mm}
\noindent
$^{\dagger}$ E-mail address: kubo@dmumpiwh.mppmu.mpg.de and
jik@hep.s.kanazawa-u.ac.jp
\\
$^{*}$On leave of absence from
 College of Liberal Arts, Kanazawa University, Japan

\newpage
\pagestyle{plain}
\section {Introduction}
In the first attempts to develop
quantum field theory, the equal-time
canonical commutation relations played an important
role
in the canonical quantization procedure.
It, however, turned out that equal-time commutators (ETC's)
are singular in general
and that
it is possible to formulate
relativistic quantum field theory without assuming the existence
 of the equal-time
canonical  commutation relations among interacting quantum fields
 \cite{streater1,blt}.
Besides, ETC's
 have lost more and more their meanings
for phenomenological applications in renormalizable
quantum field theories.

Nevertheless, it is certainly wrong to underestimate the
phenomenological as well as theoretical usefulness
of ETC's. They have been a fundamental
concept in current algebra, and have
taken undoubtedly a special place
in investigating and understanding of
anomalies \footnote{See ref. \cite{jackiw2}.}.

During the mathematization of anomalies,
especially soon after the works of Stora \cite{stora1},
Zumino \cite{zumino} and
Baulieu \cite{baulieu} in connection to the chiral anomalies,
 Faddeev \cite{faddeev1} succeeded to reveal the relation
between the gauge group cohomology
 and
the anomalous Schwinger term
\cite{goto1} in the commutator algebra
of the Gau{\ss} law operators
in chiral Yang-Mills theories.
Based on the cohomological technique, he
predicted an explicit form of the anomalous
Schwinger term, and later
many concrete calculations \cite{j} verified
his mathematical result \cite{faddeev1}.
In the course of these explicit calculations,
one re-discovered \cite{jj,rothe}
the old result \cite{buccella,johnson,brandt} that various ETC's may
 violate the Jacobi-identity, which rose
the question of whether there
exist three-cocycle anomalies, i.e.,
Jacobi-identity violating, anomalous equal-time
commutation relations among the generators
of a symmetry transformation.
Although some applications of three-cocycles
were found in quantum mechanics  \cite{jwg}, that question
in field theory has remained unanswered \cite{jackiw2}.

An anomaly has to satisfy a certain
algebraic property, the Wess-Zumino
consistency conditions \cite{wess1}. It were
the Wess-Zumino consistency conditions on the chiral
anomalies, because of which
the so-called descent equations \cite{stora1,zumino,baulieu}, discovered
as one of consequences from
the mathematization of the chiral
anomalies mentioned above, have an application in physics.
The Wess-Zumino consistency conditions (which
can be rigorously established in renormalization
theory \cite{becchi3} \footnote{See also refs.
\cite{piguet1}, and references therein. For
renormaliation of descent equations,
see ref. \cite{piguet2}.}
by applying the action principle \cite{l} and the normal product
algorithm \cite{zi1,zi2} )
have been a basic tool
in investigating the algebraic properties of different types of
anomalies\footnote{See, for instance, ref. \cite{green2,dra}}.

The Jacobi identity of ETC's can also be used
to derive the Wess-Zumino consistency conditions
on anomalous Schwinger terms, and many applications of this idea
have been reported in recent years
\cite{wh}--\cite{fujiwara4}.
In particular, it has been shown
\cite{pm,fujiwara3} that the Jacobi identity for
the BRST algebra in a BRST quantized theory
leads to the Hamiltonian descent equations. It seems that
the algebraic approach to anomalies based on
the Jacobi identity of ETC's
is  an
alternative way to investigate the algebraic structure of anomalies.
Unfortunately,
this approach suffers from the essential disadvantage,
because ETC's do not always exist, as we emphasized
at the beginning.
But this general remark does not necessarily
prevent us from dealing with
ETC's. Before applying this idea, we have to check whether
the ETC's involved in verifying an Jacobi identity
do really exist or not, and if not , at which order
in perturbation theory they start to be ill-defined.
Once their existence is established,
even in lower orders in perturbation theory,  the Jacobi identity
of ETC's can play a powerful tool to study
the algebraic nature of anomalies, as experienced in many examples
\cite{wh}--\cite{fujiwara4}.

Our concern in this paper is as follows:
 \newline
\noindent
(i) How to compute perturbatively ETC's in a renormalizable
field theory, \newline
(ii) under what conditions they  can be given
 a well-defined meaning, and
\newline
(iii) whether the Jacobi identity
of ETC's is satisfied.
\newline
Obviously, these three points are closely related, and there
are partially answers to them. In particular,
the Bjorken-Johnson-Low (BJL) method \cite{bjl,johnson} to
compute ETC's has been used over more than two decades.
In many applications of the BJL method, one encountered
``divergent'' ETC's \cite{johnson}. The appearance of those divergent
ETC's certainly reflects the
above mentioned fact that ETC's are singular in general.
However, it has not been carefully investigated whether
there is one-to-one correspondence between the appearance
of a divergent ETC in the  BJL method and the non-existence
of the ETC.

In section 2, we will begin by formally discussing the singular
nature of the equal-time limit, and then we will
investigate its convergence in momentum space by considering the
BJL method carefully
and introducing the notion of the effective dimension $d_{\rm eff}$. It
will turn out that the equal-time limit vanishes for amplitudes
with $d_{\rm eff} \leq -2$ and is finite for tree-like
amplitudes (those without nontrivial
discontinuity) with $d_{\rm eff} =-1$. Amplitudes with
$d_{\rm eff} \geq 1$ yield divergent ETC's unless
there is some unexpected cancellation mechanism.
We will then consider double ETC's and the Jacobi identity of ETC's
in section 3, and will find that the double ETC's
have to satisfy for their existence
similar power counting rules as the single ETC's.
It will be shown that, if the convergence condition
is satisfied for the ETC's involved in verifying an Jacobi identity,
the Jacobi identity is automatically satisfied.
It should be emphasized that the ECT's defined in terms of the
equal-time limit is generally deformed in that they do not
always satisfy the product rule.

Under those circumstances, it is still possible to use ETC's
to analyze quantum symmetries and their
anomalies, especially BRST symmetries and
anomalies. This is because ``$0$'' of an ETC can be shown to all orders
(we have to show that $d_{\rm eff} \leq -2$) and  the
the diagrams that are influenced by an anomaly
and yield a nontrivial Schwinger term to ``$0$'',
can behave like tree diagrams
in the lowest order.
In section 4,
we will apply the power counting rules
in  the Yang-Mills theory to
investigate the existence of the ETC's of the BRST current
$J_{\mu}$.
We will find that
  $[\, J_{\mu}(x)\,,\,J_{\nu}(0)\,]_{\rm ETC}$
only for $\mu=\nu=0$ vanishes
to all orders in perturbation theory
if the theory is free from the chiral anomaly so that we conclude
$[\, Q\,,\,Q\,]=0$,
where $Q$ is the BRST charge. (For other components, they are
power-counting divergent.)
We will also find that
in anomalous Yang-Mills theories the $O(\hbar^2)$ term of
$[\, Q\,,\,Q\,]$ is finite and the $O(\hbar^4)$ term of
$[\,	Q\,,\,[\, Q\,,\,Q\,]\,]$ vanishes,
but they are ill-defined in higher orders. This justifies the
assumption of ref. \cite{fujiwara3} from which a set of various
consistency conditions on the anomalous Schwinger terms
in chiral Yang-Mills theories has been derived.
We understand this way why the explicit computations
of the algebra of the Gau{\ss} law operators in the one-loop order
\cite{j} have yielded finite results.

Finally,
we would like to emphasize that there are no fundamental
principles that justify  renormalization or
regularization of ETC's.
 A theory with
ill-defined ETC's is not necessarily sick.
This is so for the ultraviolet as well as infrared
singularities. Our power counting rules concern
the ultraviolet behaviors only, and so ETC's
which satisfy our convergence condition might still suffer
from infrared singularities. The problem of these
infrared singularities is beyond the scope of the
present paper, and should be carefully investigated
elsewhere.

\section{Convergence criterion
for the equal-time limit}

\subsection{Singularities in the equal-time limit}
We consider a renormalizable theory, and assume
that  the theory is renormalized in
some renormalization scheme, e.g., the Bogoliubov-Parasiuk-Hepp-
Zimmermann
(BPHZ) scheme \cite{bphz}, the minimal subtraction
(MS) scheme \cite{tv,ms,breitenlohner1}, etc.
For definitenes, we work in
$D=4$ dimensions \footnote{$x^{\mu}~(\mu =0,1,2,3)$
 are coordinates, and
we employ the metric convention $(+,-,-,-)$.}. $\phi_{A}(x)$ stands for a
local,
renormalized, elementary Heisenberg field,
and $A$ includes Lorentz and also those
corresponding to internal degrees of
freedom\footnote{Its Grassman parity is
assumed to be even for simplicity.}.

The time-ordered product, T-product,
of $\phi$'s
\be
& &T\,\phi_{A_{1}}(x_{1})\phi_{A_{2}}(x_{2})
\cdots\phi_{A_{N}}(x_{N}) \nn\\
&\equiv&\phi_{A_{j_{1}}}(x_{j_{1}})
\phi_{A_{j_{2}}}(x_{j_{2}})
\cdots\phi_{A_{j_{N}}}(x_{j_{N}})~~\mbox{for}~~
x^{0}_{j_{1}}>x^{0}_{j_{2}}>\cdots >x^{0}_{j_{N}}
\ee
is defined in terms of the Green's function
$<\alpha|\,T\,\phi_{A_{1}}(x_{1})\phi_{A_{2}}(x_{2})
\cdots\phi_{A_{N}}(x_{N})\,|\beta>$,
which
is given a well-defined meaning as
a temperate distribution by assumption.
Similarly, composite operators
are well-defined in perturbation theory:
In the BPHZ renormalization scheme, they are defined in
terms of Zimmermann's
normal products \cite{zi1,zi2}, which can be appropriately extended for
MS scheme \cite{collins1}
\footnote{See also references cited
in ref. \cite{s}.}. As in the case of
T-products, their precise meaning in perturbation theory is given by
Green's  functions containing
those composite operators.

 Let denote
${\cal A}(x)$ and ${\cal B}(y)$
renormalized, local operators, elementary
or composite. We assume that the theory is invariant under Lorentz
transformations as well as translations and rotations, and
recall the definition of the T-product of
$ {\cal A}(x)$ and ${\cal B}(y)$ to
consider the equal-time limit,
\be
\lim_{x^0\to + 0}\,\{\,T\,
{\cal A}(\vec{x},x^0)\,{\cal B}(0)-
T\,{\cal A}({\vec x},-x^0)\,{\cal B}(0)\,\}~.
\ee
If the limit
has a well-defined meaning, we may define an ETC of
$ {\cal A}(x)$ and ${\cal B}(y)$.
Note that because of
local commutativity,
\be
 [\, {\cal A}(x)\,,\,{\cal B}(y)\, ] &=&0~~\mbox{for}~~
(x-y)^2<0  ~,
\ee
the expressions (2) are different from zero
only in the region $\vec{x}\simeq \vec{0}$.
Therefore, it may be convenient to apply the
operator product expansion method \cite{wilson1,zi3}
to consider the equal-time limits.

As known, the Wilson coefficients appearing in the
operator expansion method are
temperate distributions in general and so
the equal-limits do not always have a sensible
meaning.
Therefore,  in order
for the ETC to exist, we require that the smeared expression,
\be
\hat{G}(x^0) &=&\left( \begin{array}{c}
G(x^0)~\mbox{for}~~x^0 >0\\
-G(-x^0)~\mbox{for}~~x^0 <0
\end{array}\right)~,\\
G(x^0) &=& \int d^3 {\vec x}\,\chi ({\vec x})
<\alpha|\{\,T\,
{\cal A}(\vec{x},x^0)\,{\cal B}(0)-
T\,{\cal A}({\vec x},-x^0)\,{\cal B}(0)\,\}|\beta>~,\nn
\ee
is a function of $x^0$ and has an well-defined equal-time limit,
where $<\alpha|$
and $|\beta>$ denote arbitrary states \footnote{This is
the most conservative standpoint on the
equal-time limits. See ref. \cite{brandt} for a
more relaxed treatment of the equal-time limits.}.

There are two types of singularities in renormalizable
field theories that
prevent
 $\lim_{x^0 \to 0}\,\hat{G}(x^0)$ to
be well-defined \footnote{We do not consider infrared singularities
throughout this paper as already announced.}.
The one is harmless, and is related to the
 ambiguity of Green's
functions at the same
points.  Because of this arbitrariness, it is always possible
to add to two-point Green' functions
a quasi-local
distribution of the form \cite{bs}, $
{\cal P}(\partial/\partial x)\, \delta^4 (x-y)$,
where ${\cal P}(\partial/\partial x)$ is some polynomial
in $\partial/\partial x$ with constant coefficients.
Therefore, $G(x^0)$
may involve the singularities like $\delta (x^0=0)$.
 The arbitrariness mentioned here corresponds to
different choices of adding local counterterms.
Therefore, the $\delta$-singularities can be canceled
by local counterterms and are harmless.

The real singular nature in the $x^0\to 0$ limit originates from the
fact that the quantum fields
smeared in the spatial coordinates only
(as we have done above)
may still suffer from some singularities
\cite{streater1,blt} \footnote{In addition to this problem, one
could encounter another unwelcome situation that there exist
some inequivalent representations of the canonical commutation
relations, which may
be related  to Haag's theorem \cite{blt}}.
These singularity  has
a non-local nature like
\be
\lim_{x^0 \to 0}\,\hat{G}(x^0) &\sim& \frac{\ln ^p x^0 }{(x^0)^r}~.
\ee
in the equal-time limit.

The absence of
such singularities can not be ensured
in general. But
there are many examples in lower orders in
perturbation theory that suggest the existence of various
equal-time commutators
in lower orders in perturbation theory. Those computations have
been performed in
momentum space by using
the BJL method \footnote{It is
certainly possible to compute the ETC's directly
in coordinate space. The computations might
become even more economic if one could apply the technique
of the asymptotic expansions (see, for instance, ref. \cite{s}).}.
It is therefore
appropriate
to carefully investigate the relation
between the divergences that appear
in the  Bjorken limit and the singularities above.

\subsection{Power counting rules}

We use momentum
space representation
to consider the equal-time limit (2)
\footnote{We suppress the dependence of
the external states in the amplitude.}:
\be
\tilde{G}({\vec p},x^0)& =&
+(-)\,\int_{-\infty}^{+\infty} \frac{d p^0}{(2\pi)}\,
(\,e^{-ip^0 x_0}-e^{+ip^0 x_0}\,)\,T({\vec p},p^0)~
\mbox{for}~~x^0 >(<)~ 0,\\
T({\vec p},p^0)&=&\int d^4 x\,e^{ip\cdot x}\,
<\alpha|\,T\, {\cal A}(x){\cal B}(0)\,|\beta>~,\nn
\ee
where the Fourier transform of the equal-time limit, i.e.,
\be
 \int  \frac{d^{3}{\vec p}}{(2\pi)^3}\,(\exp i{\vec p}
\cdot {\vec x})~
\lim_{x^0 \to +0}\,\tilde{G}({\vec p},x^0)~,
\ee
is exactly equal to (2).
One can convince oneself from (6) that only
the odd part of amplitude,
\be
T^{\rm odd}({\vec p},p^0)
&\equiv& \frac{1}{2}\{\,T({\vec p},p^0)-T({\vec p},-p^0)\,\}~,
\ee
contributes to  (6). Our task here is
to find a sufficient condition on the amplitude $T(p)$
for $\tilde{G}({\vec p},x^0)$ to be well-defined
in the $x^0 \to 0$ limit.

Feynman amplitude may be
 analytically continued into the complex plane.
So we assume that $T^{\rm odd}({\vec p},z)$
(which is supposed to be real in some
interval on the real axis) is an analytic function
of $z$ with possible singularities no worse than
 some discontinuities on the real
 axis and poles. We may also assume that all the external state momenta
as well as  the spatial components $p_{i}$
of $\tilde{G}({\vec p},x^0)$ are kept finite
so that all the poles in $z$ are located in a finite domain
on the complex plane, and that they do not lie on the
branch cuts.
Accordingly, we extend the $z$ integral
to the lower half plane for
the first term in the parenthesis and
 to the upper half plane for
the second term because $x^0 > 0$.
We then write the complex integrals,
respectively, as a sum of two integrals:
 \be
& &\int_{-\infty}^{+\infty} \frac{d z^0}{(2\pi)}
\,e^{-(+)iz x_0}\,T^{\rm odd}({\vec p},z)~\nn\\
&=&
\oint_{C^{+(-)}_{R}} \frac{d z^0}{(2\pi)}
\,e^{-(+)iz x_0}\,T^{\rm odd}({\vec p},z)+
\int_{L^{+(-)}_{R}} \frac{d z^0}{(2\pi)}
\,e^{-(+)iz x_0}\,T^{\rm odd}({\vec p},z)~,\nn
\ee
where the contours  $C^{+}_{R}~,~L^{+}_{R}$ are shown
in fig. 1(a).
(Recall that the contours along the real axis
must be carefully chosen so that it corresponds
 to the $\epsilon$
prescription of the Feynman propagators.)

We first consider the second integral in the $R \to \infty$ limit,
and require that
\be
\lim_{R \to \infty}\,
\int_{L^{+(-)}_{R}} \frac{d z^0}{(2\pi)}
\,e^{-\,(+)iz x_0}\,T^{\rm odd}({\vec p},z)
&=&0~.
\ee
This limit exists if
the amplitude $T^{\rm odd}({\vec p},z)$
vanishes as $|z| $ approaches infinity and $|
\partial \,T^{\rm odd}({\vec p},p_0)\,/\,\partial p_0\,|$
decreases
at most like $|p_0|^{-\delta-1}$ with $\delta \geq 0$ as $|p_0|
\to\infty$. For non-oscillating amplitudes
(which is the case in general), it means
\be
\lim_{|p_0| \to \infty}\,T^{\rm odd}({\vec p},p_0)&\leq&
K\, |p_0|^{-\delta}~,
\ee
where $K$ is some positive (real) number.

To investigate the integrals on $C^{+}_{R}$
and $C^{-}_{R}$, we further divide
the integration contours; closed contours
which do not contain the branch cuts but the poles and those
along the
branch cuts.
For the first integrals (i.e., those on the closed contours), we are
allowed to interchange the equal-time
limit and the $z$ integration. This is because the $z$ integrals
produce only a sum of functions assuming the form
like
\be
(x^0)^m\,\exp i (\,f({\vec p})x^0\,)~,~m=0,1,\cdots~.\nn
\ee
We therefore set $x^0$ equal to zero and add the contributions from the
two contour integrations to obtain
\be
-\oint_{C_R} \frac{d z}{(2\pi)}\,
\,T^{\rm odd}({\vec p},z)~,
\ee
where the contour $C_R$ is shown in fig. 1(b),
where the fad lines on the real axis indicate the branch cuts.
As for the integrals along the branch cuts, we use the formula
\be
& &\lim_{\epsilon \to +0}\,\{\,T^{\rm odd}({\vec p},p_0+i\epsilon)
-T^{\rm odd}({\vec p},p_0-i\epsilon)\,\}\nn\\
&=&2i\,\mbox{Im}\,\lim_{\epsilon\to +0}\,
T^{\rm odd}({\vec p},p_0+i\epsilon)~,\nn
\ee
and
arrive at the final expression:
\be
& &\mbox{F.T.}~\left[~\lim_{x^0\to \pm 0}\,<\alpha |\{\,T\,
{\cal A}(\vec{x},x^0)\,{\cal B}(0)-
T\,{\cal A}({\vec x},-x^0)\,{\cal B}(0)\,\}|\beta>~\right]\nn\\
&=&\lim_{x^0 \to + 0}\,\tilde{G}({\vec p},x^0)~=~
-\oint_{C_R} \frac{d z}{(2\pi)}\,
\,T^{\rm odd}({\vec p},z)\nn\\
& &+2i\,\lim_{x^0\to
+0}\,\lim_{R \to\infty}\,
\{\,\int_{I^{+}_{R}}\frac{dp_0}{2\pi}\,
e^{-ip_0 x^0}+
\int_{I^{-}_{R}}\frac{dp_0}{2\pi}\, e^{+ip_0 x^0}~\}\mbox{Im}\,
\lim_{\epsilon\to +0}\,T^{\rm odd}({\vec p},p_0+i\epsilon)\,~,
\ee
where
 $I^{\pm}_{R}$
denote the line intervals
along the cuts, and
we have assumed (9).
If  the last term
vanishes either in the $R\to\infty$ or $x^0\to +0$ limit,
we are left with the first integral, which
 is the formula derived
by Johnson and Low \cite{johnson} in 1966.

Since the first contour integral of (12) is independent
of  $R$ if $R$ is sufficiently large so that all
the poles are encircled, it is sufficient for
 the equal-time limit
of left-hand side of (12) to be defined that the last term has  an
well-defined meaning.
However, we observe that the integrals along the branch cuts
are infinite in general, even if  the Bjorken limit condition,
 \be
\lim_{p_0 \to \infty}\, [\,p_0\,\mbox{Im}\,T^{\rm odd}(p)\,)]&=&
\mbox{finite}~,
\ee
is satisfied, because
\be
\lim_{x^0 \to 0}\,\int_{\Lambda}^{\infty}
d p_0 f(p_0) \cos p_0 x^0 &=&\infty,\nn
\ee
if $f(p_0)$ behaves like $p_{0}^{-1}$ as $p_0 \to \infty$.
If the Bjorken limit (13) vanishes like $p_0^{-\delta} ~(
\delta > 0)$, we
may change the $x^0\to +0$ and $R\to\infty$ limits to obtain
\be
+2i\,\lim_{x^0\to
+0}\,\lim_{R \to\infty}\,
\{\,\int_{I^{+}_{R}}\frac{dp_0}{2\pi}+
\int_{I^{-}_{R}}\frac{dp_0}{2\pi}~\}\mbox{Im}\,\lim_{\epsilon\to +0}\,
T^{\rm odd}({\vec p},p_0+i\epsilon)~,\nn
\ee
which cancels the integral (11) along the branch cuts on $C_R$.
The integral on the arc of $C_R$ vanishes in the $R\to\infty$ limit
so that the equal-time limit in question vanishes.

Therefore, the convergence condition of
the equal-time limit  of the integrals along
the branch cuts is a very strong
condition on the  Feynman amplitude, and may be satisfied only for
tree-like amplitudes, i.e., those without nontrivial discontinuity,
in accord with the general statement
\cite{streater1,blt} that ETC's do not always exist.
As for tree-like diagrams, the convergence condition
can be directly translated to
a power counting rules for Feynman diagrams because
the large $p^0$ behavior of an amplitude
is basically the same as its large
$p^{\mu}$ behavior. Therefore, the equal-time limit (12)
exists if the (tree-like) amplitude $T^{\rm
odd}$ has a negative canonical dimension, for instance.
Clearly, not every amplitude with negative dimension
contributes, and  the ``lower limit'' has already been found above.
So we conclude that the
amplitudes having
\be
\mbox{(i)}& &
d_{\rm eff}\,[\,T(p)\,]~=~-1
\ee
can give  well-defined ETC's
if there is no nontrivial discontinuity,
and those with
\be
\mbox{(ii)}  & &
d_{\rm eff}\,[T(p)\,] ~\leq ~-2~,
\ee
do not contribute to the equal-time limit (12)
to all orders in perturbation theory, where
 the effective dimension $d_{\rm eff}$
is the dimension which one obtains form the canonical dimension
if we do not count the powers of the spatial
components $p_i$ as well
as the external state momenta that are multiplied with the amplitudes.
If $d_{\rm eff} =0$ for an amplitude, we assign $d_{\rm eff} =-1$
to the amplitude
because only the odd part $T^{\rm odd}$ contributes to (12).
For amplitudes with $d_{\rm eff}\,[T(p)\,] ~\geq ~1$,
we can not expect an well-defined equal-time limit
in general.

Note that ``$0$''
on the left-hand side of (12)
can be proven in perturbation theory and has an well-defined meaning.
This is why  BRST symmetries can be investigated in terms
of ETC's because the BRST algebra is abelian, as we will
demonstrate in section .

\section {The violation of the Jacobi identity}
There are many examples of Jacobi-identity
violating ETC's
\cite{jj}--\cite{brandt}. Its origin is of
course the singular nature of the equal-time limit. Here we would like to
investigate this problem in coordinate space as well as  momentum
space in some detail.

To discuss the Jacobi identity, we have to consider
double equal-time commutators which always involves two
independent equal-time limits. Because of the singular
nature of the equal-time limit, the order of these two limits can not be
changed in general.

\subsection{An well-known example revised}
In perturbation theory, we are mostly dealing with linear
operators, which associate by definition and so have to
satisfy the Jacobi identity
\be
[\,{\cal A}(x)\,,\,[\, {\cal B}(y)\,,\,
{\cal C}(z)\,]\,]+\mbox{cyclic permutations}&=&0~.
\ee
In fact, Green's functions are computed in perturbation
theory, regardless  of
 the order of multiplication of linear operators, whether they are
elementary or composite.
So why there are  Jacobi-identity violating
ETC's? This is the question we will address below.

We consider a double ETC,
$[\,{\cal A}(x)\, ,\,[\,
{\cal B}(y)\, , \,{\cal C}(z)\,]_{\rm ETC}\,]_{\rm ETC} $,
along with its cyclic permutations.
If
\be
J_{\rm ETC}({\cal A}(x), {\cal B}(y),{\cal C}(z))
&\equiv &[\,{\cal A}(x)\,,\,[\, {\cal B}(y)\,,\,
{\cal C}(z)\,]_{\rm ETC}\,]_{\rm ETC}+
[\,{\cal C}(z)\,,\,[\, {\cal A}(x)\,,\,
{\cal B}(y)\,]_{\rm ETC}\,]_{\rm ETC}\nn\\
&  &+[\,{\cal B}(y)\,,\,[\, {\cal C}(z)\,,\,
{\cal A}(x)\,]_{\rm ETC}\,]_{\rm ETC}
\ee
vanishes, the Jacobi identity is satisfied.
We have written all the three terms to emphasize that
the orders of the equal-time limits for three double ETC's
are different. This is why (16) does not automatically imply the Jacobi
identity of ETC's, on the one hand, and
on the other hand, it suggests that the Jacobi identity of ETC's
is satisfied if we can change the order of the
different equal-time limits.
 This is exactly the
origin of the violation of Jacobi identity, as we will
see this more explicitly in a simple  example below.

We consider an well-known
Jacobi-identy violating ETC \cite{jj,rothe},
the equal-time commutators among
the vector and axial vector currents,
$V^{\mu}(x)$ and $A^{\mu}(x)~~(\mu=0,\cdots,3)$, in the theory of a
free massless fermion field in $D=4$ dimensions, to illustrate
the observation above.

 We denote the spinor
field by $\psi$, and follow the Bjorken-Drell notation for
the gamma matrices $\gamma^{\mu}$ and
$\gamma_5$ and also the singular functions.
The currents
\be
V^{\mu}(x) &\equiv&
:\,\overline{\psi}(x)\gamma^{\mu}\psi(x)\,:~,~
A^{\mu}(x) ~\equiv~
:\,\overline{\psi}(x)\gamma_5 \gamma^{\mu}\psi(x)\,:~.\nn
\ee
are normal ordered as indicated by $~:~$, and we do not need
any other specification of the regularization
to compute commutators because
the products of the singular functions we will encounter
are well-defined distributions \cite{bs} \footnote{The calculation
based on the BJL method requires a regularization because there
are superficially divergent diagrams.}.
We then consider
\be
J_{\rm ETC}(\,A^0(x),V^i(y),V^j(z) )~, ~i,j=1,2,3~,
\ee
where $J_{\rm ETC}$ is defined in (17)
(zero of which means the Jacobi identity).

To calculate ETC's for the present case, we use
the Wick theorem to derive
\be
& &[\,:\,\overline{\psi}(x)\,\Gamma^{\alpha}\,\psi(y)\,:\,,\,
:\,\overline{\psi}(z)\,\Gamma^{\beta}\,\psi(w)\,:\,]\nn\\
&=&-:\,\overline{\psi}(z)\,\Gamma^{\beta}\,(-i)S(w-x)\,
\Gamma^{\alpha}\,\psi(y)\,:
+:\,\overline{\psi}(x)\,\Gamma^{\alpha}\,(-i)S(y-z)\,
\Gamma^{\beta}\,\psi(w)\,:\nn\\
& &-Tr\,\Gamma^{\alpha}\,S^+(y-z)\,\Gamma^{\beta}\,
S^-(w-x)+
Tr\,\Gamma^{\alpha}\,S^-(y-z)\,\Gamma^{\beta}\,
S^+(w-x)~,
\ee
where \footnote{We use the same simbol for the commutators
and anticommutators.}
\be
S^{\pm}(x-y) &=&\pm
[\,\psi^{(\pm)}(x)\,,\,\overline{\psi}^{(\mp)}(y)\,]~,~
S(x-y)~=~iS^+(x-y)-iS^-(x-y)~,\nn\\
S^{\pm}(x)&=&i\,\gamma^{\mu}\partial_{\mu}\,\Delta^{\pm}(x)~,\nn\\
\Delta^{\pm}(x)&=&
-(\frac{1}{4\pi^2})\,\frac{1}{x^2\mp i0 x^0}
{}~=~   \frac{1}{(2\pi)^3}\,\int d^4 p\,\theta(\pm)\,
\delta (p^2)\,e^{-ip\cdot x}~.\nn
\ee
In the limit, $x=y \to z=w$, we encounter the products
of the singular functions of the type $S^-(-x)\,\Gamma\,S^+(x)$
which, in contrast to the product of two propagators,
are well-defined distributions \cite{bs} as announced. For instance,
\be
\Delta^-(-x)\, \Delta^+(x) &=&\frac{-1}{8\pi}\,\int
\frac{d^4 p}{(2\pi)^4}\,\theta(p^0)\,\theta(p^2)\nn\\
&=&\frac{i}{4(2\pi)^5\,x^0}\,\int
d^3 {\vec p}\, \exp (-i|{\vec p}|x^0
+i{\vec p}\cdot{\vec x})~,\nn
\ee
from which we obtain its equal-time limit
\be
\lim_{x^0 \to 0}\,\Delta^-(-x)\, \Delta^+(x) &=&
\frac{i}{16\pi^2\,x^0}\,\delta^3({\vec x})\cdots~.\nn
\ee
{}From similar calculations we find:
\be
& &\lim_{x^0 \to 0}\,S^-(-x)\,\Gamma\,S^+(x)\nn\\
& =&\frac{-i}{96\pi^2}\,[\,
\gamma^0\,\Gamma\,\gamma^0\,(\,
\frac{6}{x_0^3}-\frac{1}{x_0}{\vec \nabla}^2+\cdots\,)
+\{\,\gamma^0\,\Gamma\,\gamma^j +
\gamma^j\,\Gamma\,\gamma^0\,\}(\,(\,
-\frac{2}{x_0^2}+{\vec \nabla}^2\,)\,\partial_{j}
+\cdots\,)\nn\\
& &+\gamma^j\,\Gamma\,\gamma^k\,(\,
-\frac{2}{x_0^3}\,\delta_{jk}
+\frac{1}{x_0}\,(\,{\vec \nabla}^2\,\delta_{jk}
+2\partial_{j}\partial_{k}\,)+\cdots\,)\,
]\,\delta^3({\vec x})~.
\ee

Using this formula, we find that,
in accord with the known results \cite{jj}, the equal-time
commutators between $  A^0(x)$ and $ V^i(y)$, and between
$V^i(y) $ and $  V^j(z)$ are
canonical, i.e., identical to $i\times$ the Poisson brackets:
\be
[\, A^0 (x)\, ,\, V^i (y) \, ]_{\rm ETC} &=& 0~,
\ee
\be
[\, V^{i} (x)\, ,\, V^{j} (y) \, ]_{\rm ETC} &=& 2i
\,\epsilon_{ijk}\,A^{k}\,\delta^3 (\vec{x}-\vec{y})~,
\ee
and that the ETC of
$ A^0(x)$ and $ A^i(y)$ is divergent:
\be
\lim_{x^0 \to y^0}\,[\, A^0(x)\, ,\, A^k(y)   \,]
&=& \frac{-i}{3\pi^2}\,[\,
\frac{1}{(x_0-y_0)^2}\,\partial^{x}_{k}-\frac{1}{2}\,
{\vec \nabla}_{x}^2\,\partial^{x}_{k}+\cdots\,]\,\delta^3
({\vec x}-{\vec y})~.
\ee
Therefore, the Jacobi identity $ J_{\rm ETC}(\,A^0(x),V^i(y),V^j(z) )
=0$
is violated.

This is a typical example in which one sees that
the order of the equal-time
limits can not be freely changed. One namely finds that
\be
\lim_{x^0\to z^0}\,
\lim_{y^0 \to z^0}[\,V^i(x)\,,\,[\, V^j(y)\,,\,
A^0(z)\,]\,]&=&0~,\nn
\ee
whereas
\be
& &\lim_{y^0\to z^0}\,
\lim_{x^0 \to y^0}\,\{\,
[\,V^i(x)\,,\,[\, V^j(y)\,,\,
A^0(z)\,]\,]+[\,V^j(y)\,,\,[\,A^0(z)\,,\,
V^i(x)\,]\,]\,\}\nn\\
& =&\lim_{y^0\to z^0}\,
\lim_{x^0 \to y^0}\,\{\,
{\rm Tr}\,[\,
\gamma^i\,S^+(x-y)\,\gamma^j\,S^-(y-z)\,\gamma^0\gamma_5\,
S^-(z-x)\nn\\
 & &+
\gamma^0\gamma_5 \,S^-(z-y)\,\gamma^j\,S^+(y-x)\,\gamma^i\,
S^-(x-z)\,]-{\rm Tr}[+\leftrightarrow -]\nn\\
& &-\gamma^j\,S^+(y-x)\,\gamma^i\,S^-(x-z)\,\gamma^0\gamma_5\,
S^-(z-y)\nn\\
 & &-
\gamma^0\gamma_5 \,S^-(z-x)\,\gamma^i\,S^+(x-y)\,\gamma^j\,
S^-(y-z)\,]-{\rm Tr}[+\leftrightarrow -]\,\}\nn\\
&=&\lim_{x^0\to z^0}\,
\lim_{y^0 \to z^0}\,\{\,
{\rm Tr}\,[\,
\gamma^j\,(-i)S(y-x)\,\gamma^i\,S^+(x-z)\,\gamma^0\gamma_5\,
S^-(z-y)\nn\\
 & &+
\gamma^0\gamma_5 \,S^+(z-x)\,\gamma^i\,(-i)S(x-y)\,\gamma^j\,
S^-(y-z)\,]-{\rm Tr}[+\leftrightarrow -]\,\}\nn\\
&=&\frac{-2}{3\pi^2}\,\epsilon_{ijk}\,[\,
\frac{1}{(y_0-z_0)^2}\partial^{(y-z)}_{k}-\frac{1}{2}
{\vec \nabla}_{(y-z)}^2\,\partial^{(y-z)}_{k}+
\frac{x_0-y_0}{y_0-z_0}\,{\vec \nabla}_{(y-z)}^2\,
\partial^{(x-y)}_{k}\nn\\
& &+\cdots\,]\,
\delta^3({\vec y}-{\vec z})\delta^3({\vec x}-{\vec y})~.
\ee
To derive (24), we have applied (19) twice and (20).
By using (21),(22) and (23), one can easily confirm that
the result above is consistent with the Jacobi
identity of the commutator
(16) \footnote{This is the origin of the
observation of ref. \cite{rothe} that the
Jacobi identity can be recovered by appropriately changing
the order of Bjorken limits.}, i.e.,
\be
\lim_{x^0\to z^0}\,\lim_{y^0 \to z^0}\,
J(\,A^0(x),V^i(y),V^j(z) )&=&0~, ~i,j=1,2,3~.\nn
\ee

The example treated here is rather simple because we
can use the Wick theorem (19) and only such products
of singular functions appear that are well-defined
distributions.
 In more realistic cases
where renormalizable interactions are present, we have to
consider $T$-products which are generally more
singular than commutators. In the next section, we will
consider the violation of the Jacobi identity of ETC's in
momentum space.

\subsection{Momentum space consideration}
Double equal-time commutators have of course an
integral representation similar to (12):
\be
& &<\alpha|\,[\,{\cal A}(x)\, ,\,[\,
{\cal B}(y)\, , \,{\cal C}(0)
\,]_{\rm ETC}\,]_{\rm ETC}\,|\beta>\nn\\
&\equiv&
\lim_{x^0 \to +0}\,\lim_{y^0 \to +0}\,
<\alpha|\,\{\,
T\,{\cal A}({\vec x},x^0)\,
{\cal B}({\vec y},y^0)\,{\cal C}(0)
-T\,{\cal A}({\vec x},x^0)\,
{\cal B}({\vec y},-y^0)\,{\cal C}(0)\nn\\
& &-T\,{\cal A}({\vec x},-x^0)\,
{\cal B}({\vec y},y^0)\,{\cal C}(0)
+T\,{\cal A}({\vec x},-x^0)\,
{\cal B}({\vec y},-y^0)\,{\cal C}(0)\,\}\,|\beta>\nn\\
&=&\int \frac{d^3 p}{(2\pi)^3}\,\int \frac{d^3 q}{(2\pi)^3}
\,e^{+i{\vec p}\cdot{\vec x}+i{\vec q}\cdot{\vec y}}\,
\oint_{C^{z}} \frac{d z}{2\pi}\,\oint_{C^{w}}\frac{d w}{2\pi}
\,T^{\rm odd}({\vec p},z,{\vec q},w)~,
\ee
where
\be
<\alpha|\,T\,{\cal A}(x)\,
{\cal B}(y)\,{\cal C}(0)\,|\beta>
&=&
\int \frac{d^4 p}{(2\pi)^4}\,\int \frac{d^4 q}{(2\pi)^4}
\,e^{-ip\cdot x-i q\cdot y}\,T(p,q)~,\nn\\
T^{\rm odd}({\vec p},p^0,{\vec q},q^0)&\equiv&
\frac{1}{4}\,\{\,T({\vec p},p^0,{\vec q},q^0)
-T({\vec p},p^0,{\vec q},-q^0)\nn\\
& &-T({\vec p},-p^0,{\vec q},q^0)
+T({\vec p},-p^0,{\vec q},-q^0)\,\}~.
\ee
 Since the equal-time limit of the integrals
along the branch cuts are either $0$ or $\infty$, we have assumed  that the
amplitude has no nontrivial discontinuity in the $z$ and $w$ planes.
Note that the $w$ integration should be first performed
in such a way that $C^{w}$ encloses all the poles of the amplitude.
If the order is reversed we will obtain
\be
<\alpha|\,[\,{\cal B}(y)\, ,\,[\,
{\cal A}(x)\, , \,{\cal C}(0)
\,]_{\rm ETC}\,]_{\rm ETC}\,|\beta>~.
\ee
The formula (25) may be easily guessed from (12), but our concern is the
question of when
the last equation of (25) really exhibits the
corresponding double ETC.

First of all, the inner ETC, $[\,
{\cal B}(y)\, , \,{\cal C}(0)\,]_{\rm ETC}$,
which basically corresponds to the $w$ integral, has to exist.
Applying  the convergence condition (14) for
the inner ETC, it means that $d_{\rm eff}$ of
 $T^{\rm odd}({\vec p},p^0,{\vec q},q^0)$
with respect to $q^0$ must be equal to $-1$.
The second limit, $x^0 \to +0$, exists if the $w$ integration
produces a function of $z$ with the effective dimension $\leq -1$ with respect
to $z$.
{}From these observations, we conclude that the
amplitudes having
\be
\mbox{(iii)}& &
d_{\rm eff}\,[\,T(p,q)\,]~=~-2
\ee
can give  well-defined double ETC's
if the amplitudes have no nontrivial discontinuity, and those with
\be
\mbox{(iv)}  & &
d_{\rm eff}\,[T(p,q)\,] ~\leq ~-3~,
\ee
do not contribute to the double equal-time limits (25), where
$d_{\rm eff}$ can be obtained from the canonical dimension if
we do not count to the dimension
the powers of the spatial
components $p_i,q_i$ as well
as the external-state momenta that are multiplied with the amplitudes.
Because of the odd nature of the amplitudes (26),
we may assign the amplitude with
$d_{\rm eff}= -1$ to $d_{\rm eff}= -2$.
For amplitudes with $d_{\rm eff}\,[T(p,q)\,] ~\geq ~0$,
we can not expect an well-defined double equal-time limit
in general.

We now would like to come to investigate whether
 the Jacobi identity is satisfied.
All the ETC's involved in verifying the Jacobi identity
should be assumed to exist (otherwise, we can not give a sensible
meaning to the violation of the Jacobi identity).
One can compute those ETC's from the same
matrix element by changing the equal-time limits,
as we have mentioned in concluding (27). Another limit we can
obtain is
\be
(t^0 \to +0)\,(s^0 \to +0)~,~~\mbox{with}~~
s^0 ~\equiv~x^0-y^0~,~
t^0~\equiv ~-y^0~,\nn
\ee
which corresponds to the double ETC
\be
<\alpha|\,[\,[\,{\cal A}(x)\, ,\,
{\cal B}(y)\,]_{\rm ETC}\, ,
\,{\cal C}(0)\,]_{\rm ETC}\,|\beta>~.
\ee
Rewriting the exponent $\exp\,[-ix^0 z -i  y^0 w]$ as
$\exp\,[-is^0 z -i t^0 v]$ with $ v~\equiv~-z-w $,
one can easily find that
\be
(30) &=&-\int \frac{d^3 k}{(2\pi)^3}\,\int \frac{d^3 p}{(2\pi)^3}
\,e^{-i{\vec k}\cdot{\vec y}+i{\vec p}
\cdot({\vec x}-{\vec y})}\nn\\
& &\times\,
\oint_{C^{v}} \frac{d v}{2\pi}\,\oint_{C^{z}}\frac{d z}{2\pi}
\,T^{\rm odd}({\vec p},z,-{\vec k}-{\vec p},-v-z)~,
{}~{\vec k}~\equiv~ -{\vec p}-{\vec q}~.\nn
\ee
Therefore, the Jacobi identity corresponds to
\be
0&=&\oint \frac{d z}{2\pi}\,
\oint\frac{d w}{2\pi}\,
\,T^{\rm odd}({\vec p},z,{\vec q},w)~+
\oint \frac{d v}{2\pi}\,
\oint\frac{d z}{2\pi}
\,\,T^{\rm odd}({\vec p},z,-{\vec k}-{\vec p},-v-z)~\nn\\
& &-
\oint \frac{d w}{2\pi}\,
\oint\frac{d z}{2\pi}\,
\,T^{\rm odd}({\vec p},z,{\vec q},w)~.
\ee
To see that (31) is indeed satisfied, we express the amplitude
in the Low representation.

We first consider the case of the single ETC (6) and write the amplitude
$T(p)$ in the Low representation:
\be
T({\vec p},p_0) &=&\frac{i}{2\pi}\,\int_{-\infty}^{
\infty} d p_0'\,
[\,\frac{\rho_{AB}({\vec p},p_0')}{p_0-p_0'+i\epsilon}-
\frac{\rho_{BA}({\vec p},p_0')}{p_0-p_0'-i\epsilon}\,]~,\nn
\ee
where
\be
\rho_{AB}(p)&=&\int d^4 x\,e^{+ip\cdot x}\,
<\alpha|\,{\cal A}(x)\,{\cal B}(0)\,|\beta>~,\nn\\~
\rho_{BA}(p)&=&\int d^4 x\,e^{+ip\cdot x}\,
<\alpha|\,{\cal B}(0)\,{\cal A}(x)\,|\beta>~.
\ee
Since we assume that the equal-time limit is finite, the
amplitude must be tree-like.  So the  spectral functions $\rho$'s are
basically $\delta$-function distributions. We may futher assume that
the singularities are
located in a finite domain of $p_0 '$ because the spatial components $p_i$ and
the external-state momenta are kept finite.
 Applying the  Johnson-Low formula (12), one finds that
\be
\lim_{x^0 \to 0}\,\tilde{G}({\vec p},x^0)& =&
-i\oint_{C_R}\, \frac{d z}{2\pi}\,z\,
\int_{-\infty}^{\infty}\,\frac{d p_0'}{2\pi}~~
\frac{[\,\rho_{AB}({\vec p},p_0')-\rho_{BA}
({\vec p},-p_0')\,]}{(z-p_0'+i\epsilon)(z+p_0'-i\epsilon)}~,
\ee
where $\tilde{G}({\vec p},x^0)$
in this case is given in (6) and $C_R$ is a circle
with radius $R$.
 Since the whole result
is supposed to be independent of (sufficiently large) $R$,
 $R$ can be so chosen that all the singularities of $\rho$'s are located
in the interval $(-R'~,~R')$ with $ R'~<~R $.
It is then obvious the $z$ and $p_{0}'$ integrations in (33) may be
changed to obtain
\be
\lim_{x^0 \to  0}\,\tilde{G}({\vec p},x^0)
&=&\int_{-\infty}^{\infty}\,\frac{d p_0'}{2\pi}
[\,\rho_{AB}({\vec p},p_0')-\rho_{BA}
({\vec p},p_0')\,]~.
\ee
Remembering the definition of $\rho$'s (32), we see that
the right-hand side of (34)
is exactly the Fourier transform of
\be
 <\alpha|\,{\cal A}({\vec x},0)\,{\cal B}(0)-
{\cal B}(0)\,{\cal A}({\vec x},0)\,|\beta>
&=&<\alpha|\,[\,{\cal A}(x)\, ,\,{\cal B}(0)\,]_{\rm ETC}\,|\beta> .
\ee
 This means
that $\tilde{G}({\vec p},x^0)$
may be regarded as a continuous function in $x^0$
with the value (35) at $x^0=0$.

So  we write the amplitude $T(p,q)$ (26)
in the Low representation as we did above:
\be
T(p,q)&=&-\,\int_{-\infty}^{\infty} \frac{d p_0'}{2\pi}\,
\int_{-\infty}^{\infty} \frac{d q_0'}{2\pi}
\,[\,\frac{\rho_{ABC}({\vec p},p_0',
{\vec q},q_0')}{(p_0-p_0'+i\epsilon)(p_0+q_0-p_0'-q_0'
+i\epsilon)}\nn\\
& &-
\frac{\rho_{ACB}({\vec p},p_0',
{\vec q},q_0')}{(p_0-p_0'+i\epsilon)(q_0-q_0'
-i\epsilon)}
+\frac{\rho_{BAC}
({\vec p},p_0',
{\vec q},q_0')}{(p_0+q_0-p_0'-q_0'
+i\epsilon)(q_0-q_0'+i\epsilon)}\nn\\
& &-
\frac{\rho_{BCA}({\vec p},p_0',
{\vec q},q_0')}{(p_0-p_0'-i\epsilon)(q_0-q_0'
+i\epsilon)}
+\frac{\rho_{CAB}
({\vec p},p_0',
{\vec q},q_0')}{(p_0+q_0-p_0'-q_0'
-i\epsilon)(q_0-q_0'-i\epsilon)}\nn\\
& &
+\frac{\rho_{CBA}({\vec p},p_0',
{\vec q},q_0')}{(p_0-p_0'-i\epsilon)(p_0+q_0-p_0'
-q_0'-i\epsilon)}\,]~,
\ee
where
\be
\rho_{ABC}(p,q)&\equiv&
\int d^4 x \,d^4 y e^{ip\cdot x+iq\cdot y}\,
<\alpha|\,{\cal A}(x)\,
{\cal B}(y)\,{\cal C}(0)\,   |\beta>~,\nn
\ee
and similarly for other $\rho$'s.
Inserting (the odd part of) this expression into (31)
and recalling the result that the order of the contour
 and the $p_0'$ and $q_0'$ integrations may be changed
(in the absence of nontrivial discontinuities),
one easily finds
that  the right-had side of (31) indeed
vanishes \footnote{One finds that
the convergence condition (28) is not satisfied for the
case of section 3.1.
The discussion above is similar to that
of ref. \cite{rothe}, but emphasize
the importance of the absence of nontrivial discontinuities
in the amplitudes.}

Therefore, we conclude that
the Jacobi identity is satisfied for tree-like amplitudes.
Such amplitudes can of course
 appear in higher orders in perturbation theory.
This is why one can use
ETC's to analyze  quantum anomalies in terms of ETC's,
as we will see
in the next section.

\section{An application: The closure of the BRST algebra in Yang-Mills
theories}
\subsection{Gauge-fixed theory}
Before we go to the quantized theory of a Yang-Mills theory,
 we stay for a while in the classical approximation,
and discuss some Poisson bracket structures in
the theory.
We begin by writing down the Lagrangian
in the Landau gauge:
\be
{\cal L} &=& -\frac{1}{4}\,
F^{a}_{\mu\nu}\,F^{a\mu\nu}
+i\overline{\psi}_{\rm L}\,\gamma^{\mu}\,
D_{\mu}\,\psi_{\rm L}
+B^a\,\partial_{\mu}A^{a\mu}
-\partial^{\mu}\overline{c}^a\,D^{ab}_{\mu}\,c^b~,\nn\\
D_{\mu} &=&\partial_{\mu}-ig\,A^{a}_{\mu}\,T^a~,
D_{\mu}^{ab}~=~
\partial_{\mu}\delta^{ab}+g\,f^{acb}\,A^{c}_{\mu}~,
\ee
where
$\psi_{\rm L}$ is a left-handed Weyl field in some
representation of the gauge group $G$ and is minimally coupled
to the gauge fields $A^{a\mu}$, and
$c^a$ ($\overline{c}^a$) are the Faddeev-Popov (anti-) ghost
fields \footnote{The generators $T^a$ are assumed to be hermitian
and satisfy $[\,T^a\,,\,T^b\,]=i\,f^{abc}\,T^c$.}.

The Lagrangian is invariant under the BRST
transformation \cite{becchi3}
\be
\delta\,A^{a}_{\mu} &=& D^{ab}_{\mu} \,c^b~, \delta \,c^a~=~
-\frac{g}{2}\,f^{abc}\,c^b\,c^c~,\nn\\
\delta\,\overline{c}^a &=&-B^a~,~\delta B^a ~=~0~.
\ee
They can be generated, at least at the level
of the Poisson brackets, by the BRST charge:
\be
\delta\,\cdot&=&-\{\,Q\,,\,\cdot\,\}_{\rm PB}~,\nn
\ee
where
\be
Q &=&\int d^3 {\vec x}\,\{\,
c^a\,\varphi^a -\frac{g}{2}\,f^{abc}\,\dot{\overline{c}}^a\,
c^b\,c^c+B^a\,D_{0}^{ab}\,c^b\,\}~,
\ee
and $\varphi^a$ are the Gau{\ss} law constraints
\be
\varphi^a &=& -D_{i}^{ab}\,E^{bi} -g\overline{\psi}_{\rm L}
\gamma^{0}\,T^a\,\psi_{\rm L}~,~E^{ai}~=~-F^{a0i}~.\nn
\ee
They satisfy the Poisson bracket algebra
\be
\{\,\varphi^a({\vec x},x^0)\,,
\,\varphi^b({\vec y},x^0)\,\}_{\rm PB}&=&
g\,f^{abc}\,\varphi^c({\vec x},x^0)\,\delta({\vec x}-{\vec y})~,\nn
\ee
which ensures the ``nilpotencey'' of $Q$ at the classical
level \footnote{We use the same symbol for the
symmetric Poisson brackets as for the antisymmetric ones.
We will do so for the commutators
and anticommutators too.}:
\be
\{\,Q\,,\,Q\}_{\rm PB} &=&0~.
\ee

For our purpose, it is more convenient,
by means of
the equation of motion, to rewrite the
BRST charge (39) as \cite{kugo}
\be
Q &=&\int d^3 {\vec x}\,J_0(x)~,~\nn\\
J_{\mu}&=&
-c^a\,\partial_{\mu}{B}^a +\frac{g}{2}\,f^{abc}\,
\partial_{\mu}{\overline{c}}^{a}\,
c^b\,c^c+B^a\,D_{\mu}^{ab}\,c^b~,
\ee
so that $Q$ is expressed as an integral of the $\mu=0$
 component of a conserved current,
the BRST current $J_{\mu}$.

\subsection{The nilpotency of $Q$}
We will show that the quantum generalization of the Poisson bracket
algebra (40), $[\,Q\,,\,Q\,]~=~0$, is satisfied to all orders in
perturbation theory if the theory is free from the chiral
anomaly\footnote{Remember that the nilpotency
of Q plays an important role to ensure unitarity
in the operator formalism  \cite{kugo}.
Here we give a perturbative proof for the existence
of a nilpotent BRST charge.}.

Feynman rules are
conventional \footnote{See, for instance,
ref. \cite{iz}.}, but we remind ourselves that
the  $A-B$ propagator,
which is expressed by the line of fig. 2, is given by
\be
<\,A_{\mu}^{a}(x)\,B^b(y)\,>&=&\int
\frac{d^4 p}{(2\pi)^4}\,e^{-ip\cdot (x-y)}\,
\frac{-p_{\mu}\,\delta^{ab}}{p^2+i\epsilon}~,\nn
\ee
in coordinate space and also that
\be
<\,B^{a}(x)\,B^b(y)\,>&=&0~.\nn
\ee
We employ the dimensional regularization of ref. \cite{tv,breitenlohner1},
and the composite operators contained in the BRST charge
and current are defined as normal products
in the MS scheme. We however would like
to emphasize that the discussion below
is regularization independent in character because
we are basically applying the power counting rules
and Ward identities of renormalized amplitudes only.

Obviously, it is sufficient to consider ETC's among
the BRST currents $J_{\mu}$, and it is also more convenient to do so
because we can
rely on the Lorentz covariance.
To begin with, we assume that the theory is free from the chiral
anomaly so that all the Ward identities are satisfied
and also the BRST current $J_{\mu}$ is conserved.
Later we will take into account
the presence of the anomaly.
So we consider
\be
[\,J_{\mu}(x)\,,\,J_{\nu}(0)\,]_{\rm ETC}~.
\ee
All the diagrams which may contribute to the ETC (42) are shown
in fig. 3. We will show that
for $\mu=\nu=0$ the effective dimension of all the diagrams
are equal to or less than two, which means,
according to the convergence condition of (45),  that the ETC
(42) identically vanishes for $\mu=\nu =0$.

\vspace{.5cm}
\noindent
{\bf Diagram (a)}
\newline
\noindent
The diagram (a) has the canoical dimension of two so that it would yield
a divergent ETC if there would be no restriction on the amplitude.
The amplitude is the Fourier transform of
\be
<0|\,T\,J_{\mu}(x)\,J_{\nu}(0)\,|c(k_a)\,,c(k_b)>~,\nn
\ee
which we denote by
 \be
\delta^{a b}\,T^{\mu\nu}(p,k_a,k_b)~,
\ee
where we have factorized the group index structure.
To investigate its large momentum behavior
(which we need for the convergence criterion),
we expand the amplitude in small external-state momenta, $k_a$ and $k_b$,
and require that at each order in the expansion
$\partial_{\mu}\,J^{\mu}=0$ be satisfied:
\be
p_{\mu}\,T^{\mu\nu}(p,k_a,k_b)&=&q_{\nu}\,T^{\mu\nu}(p,k_a,k_b)~=~0~,\\
q &= &-(\,p+k_{+}\,)~,~ k_{\pm}~=~k_{a}\pm k_{b}~.\nn
\ee
The amplitude should also respect antisymmetry
\be
T^{\mu\nu}(p,k_a,k_b)&=&-T^{\mu\nu}(p,k_b,k_a)~,
\ee
which follows from the Fermi statistics of the ghost fields.

After some algebraic calculations, one finds that
up to and including $O(k^3)$ there are eleven independent
terms, $T^{\mu\nu}_{(aN)}~(N=1,\cdots,11)$, which
are explicitly written in appendix.
All the terms  are consistent with  (44) and (45)
up to and including that order in $k$.
There are remarkable cancellations among the terms for $\mu =\nu =0$;
for instance,
\be
T^{00}_{(a1)} &=&
(k_+ k_-)\,\{\,-(p_0 p_0)\,(\,{\vec p}^{\,2}+{\vec p}\cdot{\vec k_{+}}\,)+
p_0\, k_{+0}\, {\vec p}^{\,2}+{\vec p}^{\,2}\,
 {\vec p}^{\,2}\,\}~T_{(a1)}\,(p^2),\nn
\ee
where the scalar
amplitude $T_{(a1)}(p^2) $ has dimension of $-4$ so that it behaves like
$(p_0)^{-4}$ as $p_0 \to \infty$. Therefore,
\be
T^{00}_{(a1)}&\sim& (p_0)^{-2}~~\mbox{as}~~p_0\to \infty~,\nn
\ee
so that $d_{\rm eff} \leq -2$ and so according to the
convergence condition (14) it does not contributes to the ETC (42)
for $\mu=\nu=0$.  For the other amplitudes,
i.e., $T^{00}_{aN}~(N=1,\cdots,11)$, one observes the
similar cancellations which ensures that the diagram (a)
of fig. 3 can not contributes to $[\,J_0 (x)\,
,\, J_0 (0)\,]_{\rm ETC}$.

\vspace{.5cm}
\noindent
{\bf Diagram (b)}
\newline
\noindent
This diagram corresponds to the Green function
\be
<0|\,T\,J_{\mu}(x)\,J_{\nu}(0)\,|c(k_a)\,,c(k_b)\, ,
A_{\alpha}^{c}(k)>~,\nn
\ee
and its Fourier transform
\be
f^{a b c}\,T^{\mu\nu\alpha}(p,k_a,k_b,k)
\ee
and has the canonical dimension of one so that it is potentially
dangerous.
The amplitude (44) has to satisfy
the identities like (46) and  the symmetry
\be
T^{\mu\nu}(p,k_a,k_b,k)&=&T^{\mu\nu}(p,k_b,k_a,k)~,
\ee
and also the Ward identity which follows from
\be
0&=&\delta\,<0|\,T\,J_{\mu}(x)\,J_{\nu}(0)\,
\overline{c}^{a}(u)\,\overline{c}^{b}(v)\,\overline{c}^{c}(w)|0>\nn\\
& &-<0|\,T\,J_{\mu}(x)\,J_{\nu}(0)\,B^{a}(u)\,
\overline{c}^{b}(v)\,\overline{c}^{c}(w)|0>\nn\\
& &+<0|\,T\,J_{\mu}(x)\,J_{\nu}(0)\,
\overline{c}^{a}(u)\,B^{b}(v)\,\overline{c}^{c}(w)|0>\nn\\
& &-<0|\,T\,J_{\mu}(x)\,J_{\nu}(0)\,
\overline{c}^{a}(u)\,\overline{c}^{b}(v)\,B^{c}(w)|0>~,\nn
\ee
where $\delta$ is the BRST variation (32). Expressed in terms of
$T^{\mu\nu\alpha}(p,k_a,k_b,k)$, it means that
\be
0&=&k_{\alpha}\,T^{\mu\nu\alpha}(p,k_a,k_b,k)+
k_{a \alpha}\,T^{\mu\nu\alpha}(p,k,k_b,k_{a})+
k_{b \beta}\,T^{\mu\nu\alpha}(p,k_a,k,k_{b})~.
\ee
Up to and including $O(k^2)$, there are exactly eight independent
terms, $T^{\mu\nu\alpha}_{(bN)}~(N=1,\cdots,8)$,
 that satisfy all the requirements above, and they are explicitly given in
appendix. As in the previous case, one finds that for $\mu=\nu =0$ there
are cancellations that reduce $d_{\rm eff}$ of the diagram (b)
 at least down to $-2$ so that it does not contribute to the ETC
for $\mu=\nu =0$.

\vspace{.5cm}
\noindent
{\bf Diagram (c)}
\newline
\noindent
As previously, we impose the antisymmetry of the ghost lines
and the conservation of the BRST current to restrict the form
of the amplitude in the
$p \to \infty$ limit. One finds:
\be
T^{\mu\nu~abc}(p,k_a,k_b,k_c)&\rightarrow&
[\,\delta^{ab}\,\delta^{cd}(k_a -k_b)p+
\delta^{ac}\,\delta^{bd}(k_c -k_a)p\nn\\
& &\delta^{bc}\,\delta^{ad}(k_b -k_c)p\,]\,
(\,p_{\mu}p_{\nu}-p^2\,g_{\mu\nu}\,)T_{(c)}(p^2)~,\nn
\ee
which is the lowest order expression in
the large $p$ expansion. Again one sees the cancellation
for $\mu=\nu =0$.

\vspace{.5cm}
\noindent
{\bf Diagram (d)}
\newline
\noindent
There are three independent terms as far as the group index
structure is concerned:
\be
\delta^{ab}\,\delta^{cd}\,(k_a-k_b)~,~
(\delta^{ac}\,\delta^{db}-\delta^{ad}\,\delta^{bc})\,(k_c-k_d)~,~
(\delta^{ac}\,\delta^{db}+\delta^{ad}\,\delta^{bc})\,(k_a-k_b)~,\nn
\ee
where we have suppressed the Lorentz index. If one further impose the
conservation of the BRST current, one finds that in the large
$p$ limit the amplitude must be proportional to
$ p_{\mu}p_{\nu}-p^2\,g_{\mu\nu}$. For $\mu=\nu =0$, it means the
reduction of $d_{\rm eff}$ by two.
The diagrams (e) and (g) must also contain the same factor
to satisfy the conservation of the BRST current.
So the diagrams (d), (e) and (g) do not contribute
to $[\,J_{0}(x)\,,\,J_{0}(0)\,]_{\rm ETC}$.

\vspace{.5cm}
\noindent
{\bf Diagram (f)}
\newline
\noindent
At $O(k^0)$ there are seven terms that are consistent with
$\partial_{\mu}J^{\mu}=0$:
\be
& &\{\,(\,- p_{\mu} p_{\nu} +
 p^2 g_{\mu \nu}\,)\,  p_{\gamma}\,g_{\alpha \beta}\,\} ~,\\
& &\mbox{and (49) with}~~\{\,\gamma\to\alpha,\alpha\to\beta,
\beta\to\gamma,\}~,~
\{\,\gamma\to\beta,\alpha\to\gamma,\beta\to\alpha)\,\}~,\nn\\
& &\{\,(\,2 p_{\alpha}\, p_{\mu} p_{\nu} -
 p^2 \, \,p_{\nu}\,g_{\alpha \mu} -
 p^2 \, p_{\mu}\,g_{\alpha \nu}\,)\, p_{\beta}\,p_{\gamma}\nn\\
& &  -(\,p_{\alpha} \, p_{\nu}
- p^2 \,g_{\alpha \nu} \,)\,p^2\, p_{\gamma} \,g_{\beta \mu} -(\,
p_{\alpha}\,p_{\mu}
- p^2\,g_{\alpha \mu}\,)p^2\, p_{\gamma} \,g_{\beta \nu}\,\}~,\\
& &\mbox{and (50) with}~~\{\,\gamma\to\alpha,\alpha\to\beta,
\beta\to\gamma,\}~,~
\{\,\gamma\to\beta,\alpha\to\gamma,\beta\to\alpha)\,\}~,\nn\\
& & p_{\alpha}p_{\beta}p_{\gamma}\,(\,p_{\mu}p_{\nu}
-p^2\,g_{\mu\nu}  \,)~.\nn
\ee
Since the amplitude has
the canonical dimension of $-1$, we multiply the terms above
with $p_{0}^{-6}$ to consider the $p_{0}\to \infty $ limit.
One easily finds that, for instance, the term (50) for $\mu=\nu=0$ becomes
\be
2\,{\vec p}^{\,2}\,{\vec p}^{\,2}\,p_{\gamma}&
&\mbox{for}~~\alpha=\beta=0~,~ 2\,p_0 \,p_{j}\,{\vec p}^{\,2}
\,p_{\gamma}~~ \mbox{for}~~\alpha=0,\beta=j~,\nn\\
2\,p_{0}^{2}\,p_{i}
\,p_{j}\,p_{\gamma}&
&\mbox{for}~~\alpha=i,\beta=j~,\nn
\ee
and that $d_{\rm eff}$ is reduced at least by $2$ because
we do not count the spatial components $p_i$ to the dimension.

We may summarize our findings by
concluding  that the nilpotency condition of the BRST
charge
\be
Q^2&\equiv& \frac{1}{2}[\,Q\,,\,Q\,]~=~
\frac{1}{2}\int d^3{\vec x}\, d^3{\vec y}\,
[\,J_0(x)\,,\,J_(y)\,]_{\rm ETC}~=~0
\ee
is satisfied if the BRST symmetry is intact.
 The result is of course
in accord with  expectation. But we would like to
emphasize that the ETC's defined in terms of the
equal-time limit are not canonical and
there is no compelling reason for that to be true in perturbation
theory in general:
ETC's are generally ill-defined  and there is no guaranty for
the generators of a symmetry transformation to form
a closed algebra under the equal-time commutators.

\subsection{Schwinger term for $[\,Q\,,\,Q\,]$}
Until now we have assumed that the BRST current (41) is conserved
and the various Ward identities are satisfied. In the presence of the
chiral anomaly, this is no longer the case. To
investigate the effects of the anomaly, we recall that
according to the action principle \cite{l}
the violations of conservation laws and Ward identities
manifest themselves  in certain insertions in Green's
functions. For the diagram (a) of fig. 3, for example, we have to consider
\be
<0|\,\Delta_{\mu}(x)\,J_{\nu}(0)\,|c(k_a)\,,c(k_b)>~,
\ee
where $\Delta_{\mu}$ is a local insertion and given by \cite{becchi3}
(in a geometric notation)
\be
\Delta_{\mu}dx^{\mu} &=&\frac{i}{24\pi^2}\, {\rm Tr}\,
C\,d[\,A\,dA+\frac{1}{2}\,A^3\,]~,\\
C&\equiv&-ig\,T^a\,c^a~,~A~\equiv~-ig\,A^{a\mu}\,T^a\,dx_{\mu}~,\nn
\ee
where $d$ is the exterior derivative in four-dimensions.
One easily observes that there is no tree diagram for (52).
This means that for the diagram (a) the violation of the BRST symmetry
effectively appears at earliest at $O(\hbar^2)$ because
$\Delta_{\mu}$ is of $O(\hbar)$.

{}From similar considerations one finds that only the diagrams
(c) and (d) suffer from the anomaly at $O(\hbar)$ as
 shown in fig. 4(a) and (b), that is,
the effect of the anomaly appears
at the one-loop level for those diagrams. Recalling
that their dimensions are $-1$ and $-2$, respectively, and that
they happen to be tree-like diagrams,
we conclude that the $O(\hbar^2)$ term of $[\,Q\,,\,Q\,]$
has to be finite. That is, $\Omega$ defined by
\be
[\,Q\,,\,Q\,]&=&i\hbar^2 \Omega +O(\hbar^3)~,
\ee
has  an well-defined meaning while the higher order terms are
presumably ill-defined.

The violation of the Jacobi identity may be studied
as follows.
Since $[\,J_0(x),,\,J_0(0)\,]_{\rm ETC}$
vanishes, the effective dimension of
the amplitude for
\be
<0|\,T\,J_0(x)\,J_0(y)\,J_0 (0)\,|c(k_a)\,,c(k_b)\,,c(k_c)>~
\ee
has to be equal to or less that $-3$ if there
is no anomaly.
This is because for each equal-time limit
the inner ETC is the ETC of two $J_0$'s which vanishes identically
if the BRST symmetry is intact.
In the presence of the chiral anomaly, the effective dimension
will be altered, and we have to consider the large momentum behavior of
the Fourier transform of
Green's functions of the form
\be
<0|\,T\,J_{0}(x)\,J_{0}(y)\,J_{0}(0)\,
|c(k_a)\,,c(k_b)\,,c(k_c)\,,\chi>~,
\ee
where $\chi$ denotes additional external lines to (55).
But we know from the previous investigation
that only those with two and three external gauge boson lines
can have contributions at the lowest non-trivial order in $\hbar$.
The corresponding amplitudes thus
have dimensions of $0$ and $-1$, respectively, and moreover
they are tree-like diagrams (the corresponding
amplitudes do not have
nontrivial discontinuities).
Therefore, according to the conclusion of
section  3.2, the Jacobi identity has to be
satisfied in the lowest order. That is,
\be
[\,Q\, ,\,[\,Q\,,\,Q\,]\,] &=&O(\hbar^4)~,
\ee
where the $O(\hbar^4)$ term above is probably divergent.

This justifies the assumption of ref. \cite{fujiwara3}
(at least to the lowest non-trivial order
in $\hbar$) in which
a set of various consistency conditions on the anomalous
Schwinger terms for the BRST algebra has been derived
and exhaustively solved. The basic idea there was that,
starting from  (54) and assuming (57), one derives the consistency
condition on $\Omega$, \be
\{\,Q\,,\,\Omega\,\}_{\rm PB}&=0&~.\nn
\ee
This defines a classical cohomology problem, and the solution,
 unique
up to cohomologically trivial terms, is
given by \cite{fujiwara3}
\be
\Omega &=&\frac{i}{24\pi^2}\,\int \mbox{Tr}\,
\{\, C^2(\hat{A}\hat{d}\hat{A}+\hat{d}\hat{A}\hat{A}
+\hat{A}^3 +C\hat{A}C\hat{d}\hat{A}\,\}~,\nn\\
\hat{A}&=&-ig\,T^a A^{a}_{k}\,
\hat{d}x^k~(k=1,2,3)~,\nn
\ee
where $\hat{d}$ is the exterior derivative in three-dimensions,
and $\hat{A}$ is a three-dimensional one-form.

\section{Summary}
The closure of the algebra of first-class constraints
under the Poisson brackets is the expression  of the presence
of local symmetries in the classical Hamiltonian
formalism \cite{dirac}. Since the Poisson brackets are
replaced by the  equal-time commutators (ETC's)
 in the canonical quantization
procedure,
one might expect
that the corresponding constraint operators
form the same algebra under the equal-time commutators.
However, we have seen that one can not obtain
a sensible, finite equal-time limit if
the amplitude has nontrivial discontinuities. Since higher order
amplitudes have discontinuities in general, it is unlikely
possible to exhibit the Poisson bracket algebra
in terms of ETC's in renormalizable field theories.
However, any constraints algebra, closed or open, can
be expressed as a BRST algebra under
the Poisson brackets \cite{bfv}.
Since  the algebra is abelian and ``$0$'' of an ETC
can be proven in perturbation theory, it is possible to
study quantum BRST symmetries in terms of certain
equal-time commutators.
 We have in fact shown that the nilpotency
condition on the BRST charge in Yang-Mills theories
is satisfied if the chiral anomaly is canceled.

Even if  ETC's exist, they are often ``deformed''
in the sense that
the product rule is violated so that they differ from the
canonical results. So another important  question is whether
these ETC's have the derivative property,
i.e., they are Jacobi-identity satisfying, which is
needed to form an associative algebra.
We have found that, if the
the convergence condition  for double ETC's are
satisfied, the ETC's have the derivative property.
This observation can be applied to investigate the algebraic structure of
anomalous Schwinger terms, as we have done
in anomalous Yang-Mills theories.

\vspace{2cm}
\noindent
{\bf Acknowledgment}

\noindent
I would like to thank D. Maison, K. Sibold, E. Seiler, V.A. Smirnov,
H. Watanabe and W. Zimmermann for useful discussions.

\newpage
\begin{center}{\bf\Large Appendix}
\end{center}
\noindent
The independent terms in the small external-state momenta, that are
consistent with various Ward identities, are listed.
The calculation has been performed by using the
Mathematica package ``Tracer'' \cite{tr}.
\newline
\vspace{0.5cm}
\noindent
{\bf Diagram of fig. 3 (a): $T^{\mu\nu}_{aN}$}
\noindent
\be
T^{\mu\nu}_{a1}&=& [\,-p^{2}\,  p_{\mu}\, p_{\nu} +
(pk_{+})\, p_{\mu}\, p_{\nu}  -
p^{2}\,  p_{\nu} \, k_{+ \mu} +
(p^{2})^2 \, g_{\mu \nu}\,]\,p^2\,   (k_{+}k_{-})\,T_{a1}(p^2)~,\nn\\
T^{\mu\nu}_{a2}&=&[\, p_{\mu} \,p_{\nu} -
p_{\nu}\, k_{+ \mu} +
 (p^{2})^2 \,g_{\mu \nu} +
(pk_{+})\, g_{\mu \nu}\,]\,   (pk_{-})\,T_{a2}(p^2)~,\nn\\
T^{\mu\nu}_{a3}&=&[\,-p^{2}\,p_{\mu}\,p_{\nu} +
 (pk_{+})\,p_{\mu}\, p_{\nu} -
 p^{2} \,p_{\nu}\, k_{+ \mu} -
(pk_{+}) \,p_{\mu}\, k_{+ \nu}\nn\\
& & +
 p^{2}\, k_{+ \mu}\, k_{+ \nu} +
(p^{2})^2 \,g_{\mu \nu} \,]\,(p k_{-})\,T_{a3}(p^2)~,\nn\\
T^{\mu\nu}_{a4}&=&\{\,
-p^{2}\,  (pk_{-})\, p_{\mu}\, p_{\nu} -
(pk_{+})\, p_{\mu} \,p_{\nu}\, (k_{+}k_{-}) +
 p^{2}\,  p_{\nu}\, (k_{+}k_{-})\, k_{+ \mu} \nn\\+
& &p^{2}\,  (pk_{+})\, p_{\mu}\, k_{- \nu} -
 (p^{2})^2\, k_{+ \mu}\, k_{- \nu} + (p^{2})^2\,  (pk_{-})\,
 g_{\mu \nu}\,\}\,T_{a4}(p^2)~,\nn\\
T^{\mu\nu}_{a5,6,7}&=& [\, -p_{\mu}\, p_{\nu}\, +
p^{2}\, g_{\mu \nu}\,]\,\{\, k_{+}^{2}\, (pk_{-})~,~
 (pk_{+})\,(k_{+}k_{-}) ~,~ (pk_{-})\,k_{-}^{2} \,\}\,T_{a5,6,7}(p^2)~,\nn\\
T^{\mu\nu}_{a8}&=&[\,- (pk_{+})\, p_{\nu}\, k_{+ \mu} -
 (pk_{+})\,p_{\mu}\, k_{+ \nu} +
p^{2}\,  k_{+ \mu}\, k_{+ \nu} +
(pk_{+})^2\, g_{\mu \nu}\,]\,  (pk_{-})\,T_{a8}(p^2)~,\nn\\
T^{\mu\nu}_{a9}&=&  [\, (pk_{-})\,  p_{\nu}\, k_{- \mu} -
(pk_{-})\,  p_{\mu}\, k_{- \nu} +
 p^{2}\,  k_{- \mu} \,k_{- \nu} +
(pk_{-})^2\, g_{\mu \nu}\,]\, (pk_{-})\,T_{a9}(p^2)~,\nn\\
T^{\mu\nu}_{a10}&=&[\,(pk_{+})^2 \,  p_{\mu} \,p_{\nu} -
p^{2}\,(pk_{+})\,p_{\nu}\, k_{+ \mu} -
 p^{2}\, (pk_{+})\, p_{\mu} \,k_{+ \nu} +
(p^{2})^2 \,k_{+ \mu}\, k_{+ \nu}\,]\, (pk_{-})\,T_{a10}(p^2)~,\nn \\
T^{\mu\nu}_{a11}&=&[\, (pk_{-})^2\,  p_{\mu}\, p_{\nu} - p^{2}\,
 (pk_{-})\,  p_{\nu}\, k_{- \mu} -
 p^{2}\, (pk_{-})\,  p_{\mu}\, k_{- \nu} +
(p^{2})^2\,  k_{- \mu}\, k_{- \nu}\,]\, (pk_{-})\,T_{a11}(p^2)~.\nn
\ee

\vspace{0.5cm}
\noindent
{\bf Diagram fig. 3(b):$T^{\mu\nu\alpha}_{bN}$}

\be
T^{\mu\nu\alpha}_{b1}&=& \{\,k_{\nu} p^2 p_{\alpha} p_{\mu}+[\,
p_{\alpha} \,k_{b  \nu}- p_{\nu}\, k_{a \alpha} +
p_{\alpha}\, k_{a \nu} -
p_{\nu} k_{b  \alpha}+
 k_{\nu} \,p_{\alpha}  -
 k_{\alpha}\, p_{\nu} -
 p_{\alpha}\, p_{\nu}\,]\,k_{\mu}\, p^2  \nn  \\
& &+ [\,(kp)\, k_{\alpha}  -
k^2\, p_{\alpha} -
 (kk_{a})\, p_{\alpha} -
(kk_{b})\,p_{\nu}
- (kp) \,k_{a \alpha} +
  (kp) k_{b  \alpha} \,]\, p_{\mu}\, p_{\nu}\nn\\
& &+[\, - (kp)\, k_{\nu}  -
k_{\nu}p^2+ k^2\,p_{\nu}+
(kp)\,p_{\nu} +
 (kk_{a})\,p_{\nu}
 +
(kk_{b})\,p_{\nu}\nn\\
& & - (kp)\,k_{a \nu} -
(kp)\,k_{b  \nu}\, ]\,p^2\,g_{\alpha \mu} +
[\, k_{\mu} \,p^2-
(kp)\, p_{\mu}\,]\,p^2\, g_{\alpha \nu}\,\}\,T_{b1}(p^2)~,\nn\\
T^{\mu\nu\alpha}_{b2}&=& \{\,[\,(pk_{b})\, k_{b  \alpha} -
  p_{\alpha}\, k_{b}^2 +
 (pk_{a})\,k_{b  \alpha}+
k_{\alpha}\, (pk_{a}) +
k_{\alpha}\, (pk_{b})  -
(kk_{a})\, p_{\alpha} -
(kk_{b})\, p_{\alpha} \nn  \\
& &-
 p_{\alpha}\,k_{a}^2 -
2 p_{\alpha}\, (k_{a}k_{b}) +
 (pk_{a})\, k_{a \alpha} +
(pk_{b})\, k_{a \alpha}\,]\, p_{\mu} p_{\nu}  -
[\, k_{\alpha} \, p_{\nu} k_{a \mu} -
 \, p_{\alpha} p_{\nu} k_{a \mu}\nn\\
& & -
  \, p_{\nu} k_{a \alpha} k_{a \mu} +
k_{\mu}  \, p_{\alpha} k_{a \nu} +
  \, p_{\alpha} p_{\mu} k_{a \nu} +
 \, p_{\alpha} k_{a \mu} k_{a \nu}
 -
 \, p_{\nu} k_{a \mu} k_{b  \alpha} -
 k_{\alpha}  \, p_{\nu} k_{b  \mu} -
 \, p_{\alpha} p_{\nu} k_{b  \mu} \nn  \\
& &-
  \, p_{\nu} k_{a \alpha} k_{b  \mu} +
 \, p_{\alpha} k_{a \nu} k_{b  \mu} -
  \, p_{\nu} k_{b  \alpha} k_{b  \mu} +
 k_{\mu}  \, p_{\alpha} k_{b  \nu}+
   \, p_{\alpha} p_{\mu} k_{b  \nu} +
 \, p_{\alpha} k_{a \mu} k_{b  \nu} +
  \, p_{\alpha} k_{b  \mu} k_{b  \nu}\,]\, p^2 \nn\\
& &
 +[\,(kk_{a}) \,p_{\nu} +
 (kk_{b})\, p_{\nu}  +
(pk_{a})\, p_{\nu} +
 (pk_{b})\, p_{\nu}+
p_{\nu}\, k_{a}^2 +
 2 \,p_{\nu} \,(k_{a}k_{b}) -
(kp)\,  k_{a \nu}\nn\\
& & -
 p^2\,  k_{a \nu} -
 (pk_{a})\, k_{a \nu}-
 (pk_{b})\, k_{a \nu} +
p_{\nu}\, k_{b}^2 -
 (kp) \,k_{b  \nu} -
p^2\,  k_{b  \nu}\nn\\
& &-(pk_{a})\, k_{b  \nu} -
(pk_{b})\,k_{b  \nu}\,]
\,  p^2\, g_{\alpha \mu}
+[\,- k_{\mu}  \,(pk_{a}) - k_{\mu}\,(pk_{b}) -
(pk_{a})\, p_{\mu} -(pk_{b})\, p_{\mu}\nn\\
& & + (kp)\, k_{a \mu} +
p^2\,  k_{a \mu} +  (kp) \, k_{b  \mu} +
p^2 \,  k_{b  \mu}\,] \, p^2 \, g_{\alpha \nu}\,\}\,T_{b2}(p^2)~,\nn\\
T^{\mu\nu\alpha}_{b3}&=&\{\,[\,(kp)\, k_{\alpha}  -
k^2 \,p_{\alpha}\,]\, p_{\mu}\, p_{\nu} -
[\, (kp)\, k_{\alpha}  +
k^2 \, p_{\alpha}\,]\,p^2\, g_{\mu \nu}\,\}\,T_{b3}(p^2)~,\nn\\
T^{\mu\nu\alpha}_{b4}&=&\{\,[\,-(kk_{a})\, p_{\alpha} -
(kk_{b})\, p_{\alpha}\, +
 (pk_{a})\, k_{a \alpha} +
 (pk_{b})\, k_{b  \alpha}\,]\, p_{\mu} p_{\nu} \nn  \\
& &+ [\,
 (kk_{a})p_{\alpha}  +
(kk_{b})p_{\alpha}  -
  (pk_{a})\, k_{a \alpha} -
(pk_{b}) \,k_{b  \alpha}\,]\,p^2\, g_{\mu \nu}\,\}\,T_{b4}(p^2)~,\nn\\
T^{\mu\nu\alpha}_{b5(6)}&=&\{\,[\,k_{\alpha}\, (pk_{a})  +
k_{\alpha} \,(pk_{b}) -
  p_{\alpha} \,k_{a}^2 -(+)
2 p_{\alpha} \, (k_{a}k_{b})+(-)
 (pk_{b})\,k_{a \alpha} \nn\\
& &-
 p_{\alpha}\, k_{b}^2 +(-)
 (pk_{a}) \, k_{b  \alpha}\,]\, p_{\mu} p_{\nu}
+[\,
 - k_{\alpha}\,(pk_{a})  -
 k_{\alpha}\, (pk_{b})  +
 p_{\alpha}\, k_{a}^2 \nn\\
& & +(-)
 2\, p_{\alpha}\, (k_{a}k_{b})
 -(+)
 (pk_{b})\, k_{a \alpha}  +
 p_{\alpha}\, k_{b}^2  -(+)
 (pk_{a})\, k_{b  \alpha}\,]\,p^2\, g_{\mu \nu}\,\}\,T_{b5(6)}(p^2)~,\nn\\
T^{\mu\nu\alpha}_{b7}&=&\{\,[\,-k_{\nu} \, k_{a \mu} +
k_{\mu} \, k_{a \nu} -
 k_{\nu} \, k_{b  \mu} +
k_{\mu} \, k_{b  \nu} \,]\, p_{\alpha}+
 [\,k_{\nu} \, (pk_{a})  +
 k_{\nu} \, (pk_{b}) -
 (kp) \, k_{a \nu}\nn\\
& & -
(kp) \, k_{b  \nu}\,]\, g_{\alpha \mu}+[\, -
 k_{\mu} \, (pk_{a}) -
k_{\mu} \, (pk_{b}) +
 (kp) \, k_{a \mu}  +
(kp) \, k_{b  \mu}\,]\, g_{\alpha \nu}\,\}\,T_{b7}(p^2)~,\nn\\
T^{\mu\nu\alpha}_{b8}&=&\{\,[\,-(kp)\,k_{a \alpha} +
(pk_{a}) \, k_{a \alpha} -
 (kp)\, k_{b  \alpha} +
(pk_{b})k_{b  \alpha}\,]\, p_{\mu} p_{\nu} +
 [\,(kp) \, k_{a \alpha}\nn\\
& & -
(pk_{a})\, k_{a \alpha}  +
  (kp) p^2 k_{b  \alpha} g_{\mu \nu} -
(pk_{b})\, k_{b  \alpha}\,]\,p^2\, g_{\mu \nu}\,\}\,T_{b8}(p^2)~,\nn
\ee
where $T_{a,bN}(p^2)$ are scalar functions of $p^2$.

\newpage
\begin{center}
{\bf\Large Figure Captions}
\end{center}
\vspace{1cm}
\noindent
{\bf Fig. 1} Integration contours.
\newline
\vspace{0.2cm}

\noindent
{\bf Fig. 2} $A-B$ propagator.
\newline
\vspace{0.2cm}

\noindent
{\bf Fig. 3} Diagrams that contribute to  $[Q,Q]$.
\newline
\vspace{0.2cm}

\noindent
{\bf Fig. 4} Tree level contributions of the insertion $\Delta_{\mu}$,
which is indicated by $\bullet $.

\end{document}